# Hybrid Job-Driven Scheduling for Virtual MapReduce Clusters

Ming-Chang Lee

Department of Communication Systems, Simula Research Laboratory

Martin Linges vei 25, Fornebu, 1364, Norway

mclee@simula.no

Jia-Chun Lin

Department of Informatics, University of Oslo

Gaustadallèen 23 B, Oslo, N-0373, Norway

kellylin@ifi.uio.no

Ramin Yahyapour

GWDG—Gesellschaft für wissenschaftliche

Datenverarbeitung mbH Göttingen, Göttingen, Lower Saxony, Germany

ramin.yahyapour@gwdg.de

September 10, 2018

# Hybrid Job-Driven Scheduling for Virtual MapReduce Clusters

Ming-Chang Lee, Jia-Chun Lin, and Ramin Yahyapour

**Abstract**—It is cost-efficient for a tenant with a limited budget to establish a virtual MapReduce cluster by renting multiple virtual private servers (VPSs) from a VPS provider. To provide an appropriate scheduling scheme for this type of computing environment, we propose in this paper a hybrid job-driven scheduling scheme (JoSS for short) from a tenant's perspective. JoSS provides not only job-level scheduling, but also map-task level scheduling and reduce-task level scheduling. JoSS classifies MapReduce jobs based on job scale and job type and designs an appropriate scheduling policy to schedule each class of jobs. The goal is to improve data locality for both map tasks and reduce tasks, avoid job starvation, and improve job execution performance. Two variations of JoSS are further introduced to separately achieve a better map-data locality and a faster task assignment. We conduct extensive experiments to evaluate and compare the two variations with current scheduling algorithms supported by Hadoop. The results show that the two variations outperform the other tested algorithms in terms of map-data locality, reduce-data locality, and network overhead without incurring significant overhead. In addition, the two variations are separately suitable for different MapReduce-workload scenarios and provide the best job performance among all tested algorithms.

✦

## 1 INTRODUCTION

MAPREDUCE [1] is a distributed programming model proposed by Google to process vast amount of data in a parallel manner. Due to programming-model simplicity, built-in data distribution, scalability, and fault tolerance, MapReduce and its open-source implementation called Hadoop [2] have been widely employed by many companies, including Facebook, Amazon, IBM, Twitter, and Yahoo!, to process their business data. MapReduce has also been used to solve diverse applications, such as machine learning [3], data mining [4], bioinformatics [5], social network [6], and astronomy [7]. Other MapReduce-like implementations can be found in [8], [9], [10].

MapReduce enables a programmer to define a MapReduce job as a map function and a reduce function, and provides a runtime system to divide the job into multiple map tasks and reduce tasks and perform these tasks on a MapReduce cluster in parallel. Typically, a MapReduce cluster consists of a set of commodity machines/nodes located on several racks and interconnected with each other in a local area network (LAN). In this paper, we call this a conventional MapReduce cluster. Due to the fact that building and maintaining a conventional MapReduce cluster is costly for a person/organization with a limited budget, an alternative way is to establish a virtual MapReduce cluster by either renting a MapReduce framework from a MapReduce service provider (e.g., Amazon [11]) or renting multiple virtual private servers (VPSs) from a VPS provider (e.g., Linode [12] or Future Hosting [13]). Each VPS is a virtual machine with its own operating system and disk space. Due to some reasons, such as availability issue of a datacenter or resource shortage on a popular datacenter, a tenant might rent VPSs from different datacenters operated by a same VPS provider to establish his/her virtual MapReduce cluster. In this paper, we concentrate on a virtual MapReduce cluster of this type.

For a person/organization that establishes a conventional MapReduce cluster, map-data locality (which is defined as how close a map task and its input data are [14]) in the cluster is classified into node locality, rack locality, and off-rack [15] since the person/organization is aware of the physical interconnection and placement among all nodes and all racks. However, for a tenant who establishes a virtual MapReduce cluster, the tenant only knows each VPS's IP address and each VPS's datacenter location (e.g., city name). Other information such as physical machine and rack that each VPS belongs to is unreleased by the provider. Hence, from the tenant's viewpoint, the map-data locality in his/her virtual MapReduce cluster can only be classified into the following three levels:

1) VPS-locality, which means that a map task and its input data are co-located at the same VPS.
2) Cen-locality, which means that a map task and its input are within the same datacenter, but not at the same VPS.
3) off-Cen, which means that a map task and its input are located at different datacenters.

Furthermore, reduce-data locality is rarely addressed in a conventional MapReduce cluster since reducing the distance between a reduce task and its input data coming from

- M.-C. Lee is with the Department of Computer Science, National Chiao Tung University, Taiwan. E-mail: mingchang1109@gmail.com.
- J.-C. Lin is with the Department of Informatics, University of Oslo, Norway. E-mail: kellylin1219@gmail.com.
- R. Yahyapour is with GWDG—Gesellschaft für wissenschaftliche Datenverarbeitung mbH Göttingen, Göttingen, Lower Saxony, Germany. E-mail: ramin.yahyapour@gwdg.de.

all the related map tasks in a LAN is difficult. But this is achievable in a virtual MapReduce cluster comprising multiple datacenters.

Many task scheduling algorithms have been proposed [14], [15], [16], [17], [18] to improve data locality and to shorten job turnaround time, but most of them only focus on scheduling map tasks, rather than scheduling reduce tasks. Hence, employing them in a virtual MapReduce cluster might cause a low reduce-data locality. Besides, most of current scheduling algorithms are designed to achieve the node locality and rack locality for conventional MapReduce clusters, rather than achieving the VPS-locality and Cen-locality for virtual MapReduce clusters. Consequently, adopting them in a virtual MapReduce cluster might be unable to provide a high map-data locality.

In order to provide an appropriate scheduling scheme for a tenant to achieve a high map-and-reduce data locality and improve job performance in his/her virtual MapReduce cluster, in this paper we propose a hybrid job-driven scheduling scheme (JoSS for short) by providing scheduling in three levels: job, map task, and reduce task. JoSS classifies MapReduce jobs into either large or small jobs based on each job's input size to the average datacenter scale of the virtual MapReduce cluster, and further classifies small MapReduce jobs into either map-heavy or reduce-heavy based on the ratio between each job's reduce-input size and the job's map-input size. Then JoSS uses a particular scheduling policy to schedule each class of jobs such that the corresponding network traffic generated during job execution (especially for inter-datacenter traffic) can be reduced, and the corresponding job performance can be improved. In addition, we propose two variations of JoSS, named JoSS-T and JoSS-J, to guarantee a fast task assignment and to further increase the VPS-locality, respectively.

We implement JoSS-T and JoSS-J in Hadoop-0.20.2 and conduct extensive experiments to compare them with several known scheduling algorithms supported by Hadoop, including the FIFO algorithm [1], Fair scheduling algorithm [19], and Capacity scheduling algorithm [20]. The experimental results demonstrate that both JoSS-T and JoSS-J outperform the other tested algorithms in terms of map-data locality, reduce-data locality, and network overhead without causing too much overhead, regardless of job type and scale.

The contributions of this paper are as follows.

1) We introduce JoSS to appropriately schedule Map-Reduce jobs in a virtual MapReduce cluster by addressing both map-data locality and reduce-data locality from the perspective of a tenant.

2) By classifying jobs into map-heavy and reduce-heavy jobs and designing the corresponding policies to schedule each class of jobs, JoSS increases data locality and improves job performance. Furthermore, by classifying jobs into large and small jobs and scheduling them in a round-robin fashion, JoSS avoids job starvation and improves job performance.

3) A formal proof is also provided to determine the best threshold for classifying MapReduce jobs.

4) Two variations of JoSS (i.e., JoSS-T and JoSS-J) are proposed to respectively achieve two conflicting goals: speeding up task assignment and further increasing the VPS-locality.

5) We refer to a set of MapReduce benchmarks to create two different MapReduce workloads for evaluating and comparing JoSS-T and JoSS-J with three known scheduling algorithms supported by Hadoop. Moreover, a set of metrics showing data-locality, network overhead, job performance, and load balance are used to achieve a comprehensive comparison. The results confirm that JoSS-T and JoSS-J perform well for most of the metrics.

The rest of this paper is organized as follows. Sections 2 and 3 survey MapReduce and related work, respectively. Section 4 presents the details of JoSS and the two variations. Section 5 derives the best threshold to classify map-heavy jobs and reduce-heavy jobs. In Section 6, extensive experiments are conducted and experimental results are discussed. Section 7 concludes this paper and outlines our future work.

## 2   MapReduce

A MapReduce job comprises a map function and a reduce function. The map function is applied on application-specific input data structured as a series of key-value pairs to generate intermediate key-value pairs. The reduce function merges all intermediate key-value pairs related to the same key to generate final result. In Hadoop, a MapReduce cluster consists of two masters called JobTracker [2] and Name-Node [2] and a set of slaves. JobTracker coordinates and schedules the execution of MapReduce jobs, whereas Name-Node manages the distributed filesystem namespace of the cluster. Each slave provides its computation resource to execute tasks and its storage capacity to hold data. Each slave has a limited number of map slots and reduce slots to execute map tasks and reduce tasks, respectively.

Before submitting a MapReduce job $J$ to process a data file $D$, a user needs to upload $D$ to the distributed filesystem of a MapReduce cluster. The file $D$ will be divided into fixed-size blocks (e.g., 64 MB in Hadoop [16]), and each block will be replicated and randomly stored in several slaves based on available storage space. The execution of $J$ comprises three phases: map, shuffle, and reduce. During the map phase, each map task of $J$ is assigned to a slave (we call it mapper) to process a block of $D$. If the mapper can retrieve the block from its local disk, it immediately executes the map task. Otherwise, it needs to retrieve the block from another slave, implying that the network traffic might increase and the execution of $J$ may prolong. When a mapper completes its map task, the shuffle phase starts in which the intermediate data generated by the mapper is partitioned and transmitted to each slave that is assigned to run the reduce task of $J$ (we call it reducer). After the shuffle phase ends, the reduce phase starts in which each reducer executes the user-defined reduce function to generate the final result.

## 3   Related Work

The FIFO algorithm [1] is a default scheduling algorithm provided by Hadoop MRv1. It follows a strict job submission order to schedule each map task of a job and meanwhile attempts to schedule a map task to an idle node that is close to the corresponding map-input block. However, the FIFO algorithm only focuses on map-task scheduling, rather than reduce-task scheduling. Hence, when FIFO is adopted in a virtual MapReduce cluster, its low reduce-data locality might cause a long job turnaround time. Besides, FIFO is designed to achieve node locality and rack locality in conventional MapReduce clusters, rather than achieving the VPS-locality and Cen-locality in a virtual MapReduce cluster. Consequently, the map-data locality of FIFO might be low in a virtual MapReduce cluster.

In addition to the FIFO algorithm, Hadoop also provides the fair scheduling algorithm [19] and the capacity scheduling algorithm [20]. The former is proposed by Facebook to fairly assign computation resources to jobs such that all jobs obtain an equal share of resources over time. The latter, introduced by Yahoo!, also allows multiple users to share a MapReduce cluster. It supports multiple queues and allocates a fraction of a cluster's computation resources to each queue, i.e., all jobs submitted to a queue can only access to the resource allocated to the queue. Similar to these two algorithms, JoSS allows multiple jobs to simultaneously share the computation resource of a virtual MapReduce cluster. But different from the two algorithms, JoSS further provides reduce-task scheduling to improve job performance.

There have been many studies [14], [15], [17], [18], [21], [35] on MapReduce task scheduling. Zaharia et al. [17] presented the delay scheduling algorithm to improve data locality by following the FIFO algorithm but relaxing the strict FIFO job order. If the scheduling heuristic cannot schedule a local map task, it postpones the execution of the corresponding job and searches for another local map task from pending jobs. A similar but improved approach is further introduced in [15]. However, similar to FIFO, this approach did not provide reduce-task scheduling. Jin et al. [18] proposed the BAlance-Reduce (BAR) algorithm, which produces an initial task allocation for all map tasks of a job and then takes network state and cluster workload into consideration to interactively adjust the task allocation to reduce job turnaround time. In order to simplify BAR, the authors assumed that all local map tasks spend identical execution time. But this assumption is not realistic since the map-task execution time fluctuates even though when the processed input size is the same. Besides, reduce-task scheduling was not addressed by BAR.

Tian et al. [35] proposed a MapReduce workload prediction mechanism to classify MapReduce workloads into three categories based on their CPU and I/O utilizations and then proposed a Triple-Queue Scheduler to improve the usage of both CPU and disk I/O resources under heterogeneous workloads. Guo [14] presented an optimal map-task scheduling algorithm, which converts a task assignment problem into a Linear Sum Assignment Problem so as to find the optimal assignment. Nevertheless, applying this algorithm to real-world MapReduce clusters needs to carefully determine an appropriate time point to conduct the algorithm since slaves might become idle at different time points. Ehsan and Sion [21] introduced a co-scheduler called LiPS, which utilizes linear programming to simultaneously co-schedule map-input data and map tasks to nodes such that dollar cost can be minimized. But their assumption, i.e., MapReduce jobs and their input data are submitted together, might increase job turnaround time since replicating the data to the distributed filesystem of the cluster needs to take a while. Polo et al. [36] introduced a task scheduler to dynamically predict the performance of concurrent MapReduce jobs and adjust the resource allocation for the jobs. The goal is to allow MapReduce jobs to meet their performance objectives without over-provisioning of physical resources.

Some other studies aim to enhance the performance of MapReduce in a cloud environment. Palanisamy et al. [22] presented a MapReduce resource allocation system called Purlieus, which enables a cloud provider to place MapReduce input data to appropriate physical machines and then place VMs to the physical machines so as to provide both map locality and reduce locality. Different from Purlieus, JoSS presented in this paper is designed from the perspective of a tenant who rents VPSs from a VPS provider to build a virtual MapReduce cluster, rather than from the perspective of a cloud provider. Park et al. [23] introduced a locality-aware dynamic VM reconfiguration technique for virtual clusters running the Hadoop platform by dynamically changing the computing resource of a VM to maximize the data locality of map tasks. Bu et al. [24] proposed a task scheduling strategy called ILA to mitigate interference between virtual machines and meanwhile preserve MapReduce task data locality. Similar to [22], the schemes proposed in [23] and [24] were designed from the viewpoint of a cloud provider since the data locality in all layers including node locality, rack locality, and off-rack are clear to the provider. However, in a virtual MapReduce cluster considered in this study, a tenant does not know all of the abovementioned data-locality levels.

## 4 THE PROPOSED SCHEME

In this section, we describe how JoSS schedules MapReduce jobs in a virtual MapReduce cluster consisting of $k$ datacenters, $k > 1$. Let $cen_c$ be the $c$th datacenter supporting the composition of the virtual MapReduce cluster, $c = 1, 2, \ldots, k$. Let $N_{VPS,c}$ be the number of VPSs provided by $cen_c$, $N_{VPS,c} > 1$. Let $VPS_{c,\ell}$ be the $\ell$th VPS provided by $cen_c$, $\ell = 1, 2, \ldots, N_{VPS,c}$. Assume that each VPS has only one map slot and one reduce slot, i.e., at most one map task and one reduce task can be performed by a VPS simultaneously. For each datacenter $cen_c$, JoSS maintains two permanent queues, denoted by $MQ_{c,0}$ and $RQ_{c,0}$, to respectively put the map tasks and the reduce tasks that are scheduled to be executed by VPSs at $cen_c$.

Let $J$ be a MapReduce job submitted by a user, and $D$ is the input data processed by $J$. Based on the predefined block size $S$, $D$ will be divided into $m$ blocks $B_1, B_2, \ldots, B_m$ where $m = \lceil \frac{|D|}{S} \rceil$. Let $B_i$ be the $i$-th block of $D$, $i = 1, 2, \ldots, m$. According to the total number of the blocks, $J$ is divided into the same number of map tasks. Let $M_i$ be the $i$th map task that processes $B_i$, $i = 1, 2, \ldots, m$. Let $r$ be the number of reduce tasks of $J$, and let $R_j$ be the $j$th reduce task of $J$ where $j = 1, 2, \ldots, r$ and $r \geq 1$.

In the following, a VPS performing a map task is called a mapper, whereas a VPS running a reduce task is called a reducer.

### 4.1 Job Classification

Before introducing the algorithm of JoSS, we first describe how JoSS classifies jobs and schedules each class of jobs.

TABLE 1
The Occurrence Frequencies of Top 10 Words
in One Web Document

| Word | Occurrence | Percent | Rank |
|------|-----------|---------|------|
| /> | 7,796 | 3.99% | 1 |
| \<contributor\> | 6,294 | 3.22% | 2 |
| \</contributor\> | 6,294 | 3.22% | 2 |
| \</page\> | 6,294 | 3.22% | 2 |
| \</revision\> | 6,294 | 3.22% | 2 |
| \<format\>text/x-wiki\</format\> | 6,294 | 3.22% | 2 |
| \<text\> | 6,294 | 3.22% | 2 |
| \<revision\> | 6,294 | 3.22% | 2 |
| \<model\>wikitext\</model\> | 6,294 | 3.22% | 2 |
| \<page\> | 6,294 | 3.22% | 2 |

TABLE 2
The Analysis of Word Length in One Web Document

| | |
|---|---|
| Average Word length | 22.04 |
| Std. | 12.73 |
| Max word length | 114 |
| Min word length | 1 |
| Mode | 2 |

Let $S_{reduce}$ and $S_{map}$ be the total reduce-input size and the total map-input size of $J$, respectively. Based on the ratio of $S_{reduce}$ over $S_{map}$, $J$ can be classified into either a reduce-heavy job or a map-heavy job. If $J$ satisfies Eq. (1), implying that the network overhead is dominated by $J$'s reduce-input data, then $J$ is classified as a reduce-heavy job (RH job for short). Otherwise, $J$ is classified as a map-heavy job (MH job for short). Note that $td$ is a threshold to determine the classification, $td \geq 0$. The best value of $td$ will be derived in Section 5.

$$\frac{S_{reduce}}{S_{map}} > td \qquad (1)$$

In fact, $S_{map} = \sum_{i=1}^{m}|B_i|$ where $|B_i|$ is the size of $B_i$, and $S_{reduce} = \sum_{i=1}^{m}(|B_i| \cdot FP_i)$ where $FP_i$ is the filtering percentage of $B_i$ showing the ratio of $M_i$'s map-output size over $M_i$'s map-input size, $FP_i \geq 0$ [25] [26].

In order to reduce Eq. (1) and the above classification, we chose six MapReduce benchmarks: Word-Count, Grep, Inverted-Index, Sequence-Count, Self-Join, and Term-Vector from PUMA [33] to conduct two experiments. The purpose is to study the difference among the filtering-percentage values of all map tasks of a MapReduce job. In the first experiment, we randomly selected 17 web documents from the Wikipedia dataset [30] to be the input of each benchmark. However, in the second experiment, we randomly chose ten different TXT files from [34] to be the input of each benchmark. The motivation behind these two experiments is to see whether different types of input data influence the filtering-percentage values of map tasks of a MapReduce job or not. Hence, we did not modify the six MapReduce benchmarks to suit different types of input data in our experiments.

In the first experiment, the sizes of these 17 web documents are 3.5, 5.8, 11, 35, 52, 63.5, 88.5, 172, 242, 311, 413, 546, 595, 827, 1074, 1286, and 1442 MB. Tables 1 and 2 list the occurrence frequencies of top 10 words and the analysis of word length in one of these documents, respectively. Note that the analyses of all the 17 documents are similar to Tables 1 and 2, thus we do not show them here to save space. In the first experiment, each web document was partitioned based on the block size of 128 MB. Hence, each benchmark processed 56 blocks in total, i.e., 56 map tasks were correspondingly generated and executed for each benchmark. It also implies that we could obtain 56 filtering-percentage values after each benchmark completes.

Note that we performed the Grep benchmark three times to individually search for two common patterns (e.g., a and the) and one uncommon pattern (e.g., mapreduce) in these 17 files. The purpose is to see how different input patterns/keywords impact the filtering-percentage value of the Grep benchmark.

Fig. 1 shows the average filtering-percentage values of all tested benchmarks on these 17 files. We can see that each benchmark has its own average filtering-percentage value, and all benchmarks (except for Grep) had a standard deviation of less than 0.037. Therefore, for most tested benchmarks, it is acceptable to use their average filtering-percentage values to represent the filtering-percentage values of all their map tasks.

Although Grep has a higher standard deviation because of the inputted patterns, its filtering-percentage value is at most one since its intermediate data is at most as large as its input data. Hence, any Grep or Grep-like job will always be classified as a MH job based on Eq. (3) and the best value of $td$ that will be both described later.

In addition to the above experiment, we also executed all the benchmarks on the same 17 files by setting block size into 64 MB. The corresponding filtering-percentage results are very close to Fig. 1, so they are not presented in this paper in order to save paper space. Based on our experiment results, we conclude that block size is not a key factor in determining the filtering-percentage value of a map task.

In the second experiment, the sizes of the ten TXT files are 163, 262, 292, 394, 462, 675, 702, 916, 1005, and 1057 KB. Tables 3 and 4 list the occurrence frequencies of top 10 words and the word-length analysis in one of these files, respectively. Note that the analyses of the rest files are similar to Tables 3 and 4, so again we do not show them here to save space. The block size remains the same (i.e., 128 MB).

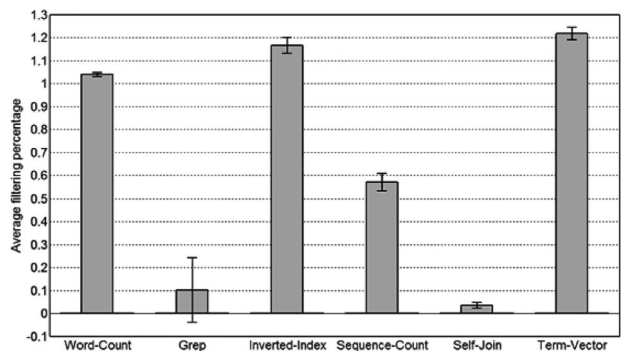

Fig. 1. The average filtering-percentage values of various mapreduce benchmarks on the 17 web documents.

TABLE 3
The Top 10 Word Frequencies in One TXT File

| Word | Occurrence | Percent | Rank |
|------|-----------|---------|------|
| the | 9,937 | 9% | 1 |
| to | 3,709 | 3% | 2 |
| and | 3,689 | 3% | 3 |
| of | 3,504 | 3% | 4 |
| a | 3,434 | 3% | 5 |
| in | 2,619 | 2% | 6 |
| I | 2,214 | 2% | 7 |
| you | 2,048 | 2% | 8 |
| it | 1,370 | 2% | 9 |
| on | 1,050 | 1% | 10 |

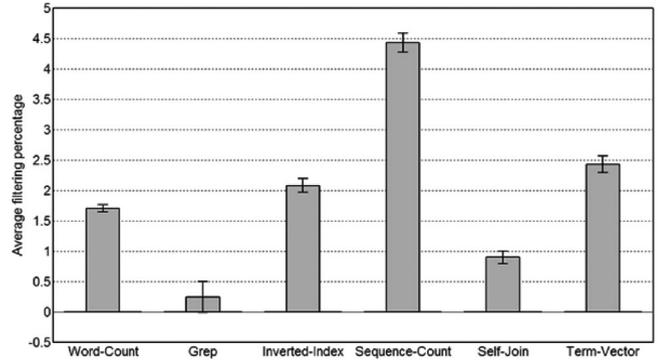

Fig. 2. The average filtering-percentage values of various MapReduce benchmarks on the 10 TXT files.

Similar to the first experiment, we also executed the Grep benchmark for three times to individually search for patterns 'a', 'the', and 'book'. Fig. 2 illustrates the average filtering-percentage values of all tested benchmarks on the ten files. It is clear that the average filtering-percentage value of each benchmark in Fig. 2 is different from that in Fig. 1, implying that the type of input-data processed by a MapReduce job has a significant impact on the corresponding filtering-percentage value. The key reason is that the numbers of whitespace characters in a web document is different from that of a non-web document. A web document usually contains a lot of whitespace characters to form all markups, but a non-web document usually does not have so many whitespace characters.

Nevertheless, Fig. 2 shows that all benchmarks (except for Grep) had a standard deviation of less than 0.15. Based on the results shown in Figs. 1 and 2, we can conclude that as long as a MapReduce job processes a same type of input data, the filtering-percentage values of all the map tasks will be similar and the standard deviation will be negligible as compared with the corresponding average filtering-percentage value. Hence, using the average filtering-percentage value to represent the filtering-percentage values of all the map tasks is acceptable. This phenomenon holds for most tested MapReduce benchmarks. Hence, in this paper, we use the average filtering-percentage value of a job on a particular input-data type to replace the filtering-percentage value of each map task of the job. In other words, we use $FP_J$ to substitute $FP_i$ where $FP_J$ is the average filtering-percentage value of $J$ and $i = 1, 2, \ldots, m$. By doing so, Eq. (1) can be reduced as

$$\frac{S_{reduce}}{S_{map}} = \frac{\sum_{i=1}^{m}(|B_i| \cdot FP_i)}{\sum_{i=1}^{m}|B_i|} = \frac{(\sum_{i=1}^{m}|B_i|) \cdot FP_J}{\sum_{i=1}^{m}|B_i|} = FP_J > td, \tag{2}$$

TABLE 4
The Word-Length Analysis of One TXT File

| | |
|---|---|
| Average Word length | 7.767 |
| Std. | 2.7 |
| Max word length | 29 |
| Min word length | 1 |
| Mode | 3 |

and the condition used to classify $J$ can be reduced as

$$J = \begin{cases} \text{a RH job,} & \text{if } FP_J > td \\ \text{a MH job,} & \text{else.} \end{cases} \tag{3}$$

In addition, JoSS also adopts another classification to classify $J$. It is based on the input scale of $J$ to $N_{avg\_VPS}$, which is the average datacenter scale of a virtual MapReduce cluster, i.e., $N_{avg\_VPS} = \frac{\sum_{c=1}^{k} N_{VPS,c}}{k}$. If $m \le N_{avg\_VPS}$ (implying that all map tasks of $J$ are possible to be performed by a single datacenter of the virtual MapReduce cluster simultaneously), $J$ is classified as a small job to the cluster. Otherwise, $J$ is classified as a large job to the cluster. In short, the classification rule is below.

$$J = \begin{cases} \text{a small job,} & \text{if } m \le N_{avg\_VPS} \\ \text{a large job,} & \text{else} \end{cases} \tag{4}$$

The purpose behind this classification is to prevent the VPSs at one datacenter of a small virtual MapReduce cluster from executing all map tasks of a large job by themselves since this will prolong job execution.

### 4.2 Scheduling Policies

Based on the job classifications mentioned in Section 4.1, JoSS utilizes the following three scheduling policies.

#### 4.2.1 Policy A

This policy is designed for a small RH job. If $J$ is a small RH job, it would be better that each reducer of $J$ is close to all mappers of $J$ since the reducer can more quickly retrieve its input data from all the mappers. But this also implies that all mappers of $J$ should be close to each other.

Hence, policy A works as follows. It first chooses $cen_w$, which is a datacenter having the least amount of unprocessed tasks among all the $k$ datacenters, $cen_w \in \{cen_1, cen_2, \ldots, cen_k\}$. Then it schedules all tasks of $J$ to $cen_w$ by putting $J$'s map tasks and $J$'s reduce tasks at the end of $MQ_{w,0}$ and $RQ_{w,0}$, respectively. In this way, all these tasks can be executed only by the VPSs at $cen_w$, and each reducer of $J$ can retrieve its input data from its local datacenter (i.e., reduce-data locality can be improved).

#### 4.2.2 Policy B

This policy is designed for a small MH job. If $J$ is a small MH job, it would be better that each mapper of $J$ is close to its input block, and each reducer of $J$ is close to most mappers of $J$. Hence, policy B works as follows: It schedules $J$'s map tasks based on the number of unique input blocks of $J$ held by each datacenter. If a datacenter holds more unique blocks of $J$, more map tasks of $J$ will be scheduled to the VPSs at this datacenter. The purpose is allowing each mapper of $J$ to retrieve its input block from its local datacenter. In addition, to make $J$'s reducers close to most $J$'s mappers, policy B schedules all reduce tasks of $J$ to the datacenter that holds the maximum number of $J$'s unique blocks.

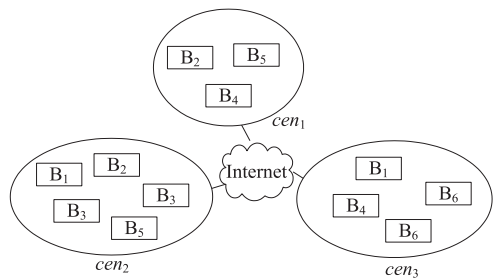

Fig. 3. An example showing the block locations of job $Y$ in a virtual MapReduce cluster comprising three datacenters.

For example, Fig. 3 illustrates the locations of all blocks of a job $Y$ over three datacenters (Note that the input file of $Y$ is fragmented into six blocks, and each block has two replicas). Since $cen_2$ holds the largest number of $Y$'s unique blocks (i.e., four), policy B will schedule four map tasks of $Y$ to $cen_2$ to process $B_1$, $B_2$, $B_3$, and $B_5$ by appending the four map tasks to the end of $MQ_{2,0}$ (Recall that $MQ_{c,0}$ is the permanent map-task queue of $cen_c$, $c = 1, 2, \ldots, k$). After that, $cen_1$ still holds one unscheduled block of $Y$ (i.e., $B_4$), and $cen_3$ still holds two unscheduled blocks of $Y$ (i.e., $B_4$ and $B_6$). Hence, policy B will schedule the remaining two map tasks of $Y$ to $cen_3$ to process $B_4$ and $B_6$ by inserting the two map tasks to the end of $MQ_{3,0}$. Finally, due to the fact that $cen_2$ holds the maximum number of unique blocks of $Y$, policy B schedules all reduce tasks of $Y$ to $cen_2$ by appending them to the end of $RQ_{2,0}$ (Recall that $RQ_{c,0}$ is the permanent reduce-task queue of $cen_c$, $c = 1, 2, \ldots, k$).

#### 4.2.3 Policy C

This policy is designed for a large job. If $J$ is a large job to a virtual MapReduce cluster, using one datacenter of the cluster to run all map tasks of $J$ might need several rounds to finish these map tasks, implying that job turnaround time will prolong. To prevent this from happening, it is better not to use a single datacenter to run all these map tasks.

Hence, as long as $J$ is a large job, JoSS utilizes policy C, which in fact uses the same strategy of policy B to schedule all tasks of $J$. However, in policy C, all the map tasks scheduled to $cen_c$ will not be put into $MQ_{c,0}$ since $MQ_{c,0}$ is reserved for only small jobs. Instead, these map tasks will be put into a new map-task queue created for $cen_c$.

Similarly, the reduce tasks of the large job scheduled to $cen_c$ will be put into a new reduce-task queue created for $cen_c$, rather than $RQ_{c,0}$. The purpose is to separate large jobs and small jobs into different queues and allow JoSS to avoid job starvation (which will be described later).

---

**The task scheduler of JoSS**

**Input:** $J$ and input-data description
**Output:** task-scheduling decision
**Procedure:**

1:   Calculate a hash value for $J$'s executable code and $J$'s input-data type;
2:   Let $H$ be a set of hash values previously generated by JoSS;
3:   **if** the hash value is not in $H$ {
4:       Append all map tasks of $J$ to the end of $MQ_{FIFO}$;
5:       Append all reduce tasks of $J$ to the end of $RQ_{FIFO}$;}
6:   **else** {
7:       **if** $J$ is a small RH job { //Use policy A.
8:           Let $cen_w$ be a datacenter having the least unprocessed
9:           tasks among $cen_1$, $cen_2$, …, $cen_k$;
10:          Append all map tasks of $J$ to the end of $MQ_{w,0}$;
11:          Append all reduce tasks of $J$ to the end of $RQ_{w,0}$; }
12:      **else**{
13:          Let $L_c$ be a set of all unique input blocks of $J$ held by $cen_c$
14:          where $c = 1, 2, …, k$;
15:          Let $\alpha = m$;   /* $m$ is the number of map tasks of $J$. */
16:          **while** $\alpha > 0$ { /* i.e., not all map tasks of $J$ are scheduled.*/
17:              Let $L_d$ be the first largest set among $L_1, L_2, …, L_k$;
18:              Let $|L_d|$ be the size of $L_d$;
19:              Let $cen_d$ be the related datacenter;
20:              **if** $J$ is a small MH job { //Use policy B
21:                  Append $|L_d|$ map tasks of $J$ to the end of $MQ_{d,0}$;}
22:              **else** { /* i.e., $J$ is a large job, so use policy C. */
23:                  Let $p$ be the total number of map-task queues in $cen_d$;
24:                  Generate a new map-task queue $MQ_{d,p+1}$;
25:                  Append $|L_d|$ map tasks of $J$ to the end of $MQ_{d,p+1}$;}
26:              **for** $c = 1$ to $k${
27:                  Delete a block from $L_c$ if the block is in $L_d$; }
28:              $\alpha = \alpha - |L_d|$;}
29:          Let $cen_e$ be a datacenter holding the largest number of
30:          unique input blocks of $J$;
31:          **if** $J$ is a small MH job { //Use policy B
32:              Append all reduce tasks of $J$ to the end of $RQ_{e,0}$; }
33:          **else** { /* i.e., $J$ is a large job, so use policy C. */
34:              Let $q$ be the total number of reduce-task queues in $cen_e$;
35:              Generate a new reduce-task queue $RQ_{e,q+1}$;
36:              Append all reduce tasks of $J$ to the end of $RQ_{e,q+1}$; }}}

Fig. 4. The algorithm of the task scheduler.

### 4.3   JoSS and its Two Variations

JoSS consists of three components: input-data classifier, task scheduler, and task assigner. The input-data classifier is designed to classify input data uploaded by a user into one of the two types: web document and non-web document. A web document refers to a file consisting of a lot of tags enclosed in angle brackets. By simply inspecting the first several sentences of a document, the input-data classifer can easily know if it is a web document or not. After the classification, the input-data classifier records the type of the input data in JoSS.

Whenever receiving a MapReduce job from a user, the task scheduler determines the type of the job and then schedules the job based on one of policies A, B, and C. The task assigner then determines how to assign a task to a VPS whenever the VPS has an idle slot.

Fig. 4 illustrates the algorithm of the task scheduler. Upon receiving $J$, the task scheduler retrieves $J$'s input-data type classified by the input-data classifier and checks whether JoSS has executed $J$ on such input-data type or not by calculating the corresponding hash value and comparing the value with $H$, where $H$ is a set of hash values previously generated and recorded by JoSS.

If the hash value is not in $H$ (see line 4), it means that JoSS does not know $J$'s average filtering-percentage value and

**Task-driven Task Assigner (TTA)**
**Input:** an idle slot of $VPS_{c,\ell}$
**Output:** a task assigned to $VPS_{c,\ell}$
**Procedure:**

```
1:   Let I_map and I_red be two indexes with the same initial value 0;
2:   while VPS_c,ℓ has an idle slot {
3:       Let N_map be the total number of map-task queues in cen_c;
4:       Let N_red be the total number of reduce-task queues in cen_c;
5:       if the slot is a map slot {
6:           if MQ_FIFO is not empty{
7:               Use FIFO to assign a map task from MQ_FIFO to VPS_c,ℓ;
8:               Remove the task from MQ_FIFO; }
9:           else {
10:              I_map = I_map mod (N_map + 1);
11:              Assign the first map task from MQ_c,I_map to VPS_c,ℓ;
12:              Remove the task from MQ_c,I_map;
13:              I_map++;}}
14:      else {  /* i.e., the idle slot is a reduce slot; */
15:          if RQ_FIFO is not empty {
16:              Assign the first reduce task from RQ_FIFO to VPS_c,ℓ;
17:              Remove the task from RQ_FIFO; }
18:          else {
19:              I_red = I_red mod (N_red + 1);
20:              Assign the first reduce task from RQ_c,I_red to VPS_c,ℓ;
21:              Remove the task from RQ_c,I_red;
22:              I_red++;}}}
```

Fig. 5. The algorithm of task-driven task assigner (TTA).

**Job-driven Task Assigner (JTA)**
**Input:** an idle slot of $VPS_{c,\ell}$
**Output:** a task assigned to $VPS_{c,\ell}$
**Procedure:**

```
1:   Let I_map and I_red be two indexes with the same initial value 0;
2:   while VPS_c,ℓ has an idle slot {
3:       Let N_map be the total number of map-task queues in cen_c;
4:       Let N_red be the total number of reduce-task queues in cen_c;
5:       if the slot is a map slot {
6:           if MQ_FIFO is not empty{
7:               Use FIFO to assign a map task from MQ_FIFO to VPS_c,ℓ;
8:               Remove the task from MQ_FIFO; }
9:           else {
10:              I_map = I_map mod (N_map + 1);
11:              Use FIFO to assign a map task from MQ_c,I_map to VPS_c,ℓ;
12:              Remove the task from MQ_c,I_map;
13:              I_map++;}}
14:      else {  /* i.e., the idle slot is a reduce slot; */
15:          if RQ_FIFO is not empty {
16:              Assign the first reduce task from RQ_FIFO to VPS_c,ℓ;
17:              Remove the task from RQ_FIFO; }
18:          else {
19:              I_red = I_red mod (N_red + 1);
20:              Assign the first reduce task from RQ_c,I_red to VPS_c,ℓ;
21:              Remove the task from RQ_c,I_red;
22:              I_red++;}}}
```

Fig. 6. The algorithm of job-driven task assigner (JTA).

$J$'s job classification. To obtain the above information, the task scheduler simply appends $J$'s all map tasks and $J$'s all reduce tasks to two queues, denoted by $MQ_{FIFO}$ and $RQ_{FIFO}$, respectively. This allows the task assigner to use the Hadoop FIFO algorithm [1] to assign these tasks to idle VPSs. Once $J$ is completed, JoSS records the corresponding hash value and average filtering-percentage value.

However, if the hash value is in $H$ (see line 7), it means that JoSS knows the average filtering-percentage value of $J$. Then the task scheduler schedules $J$ as follows: If $J$ is a small RH job, the abovementioned policy A is used to schedule the tasks of $J$ (please see lines 9 to 12). Otherwise, it means that $J$ is either a small MH job or a large job, and the task scheduler uses lines 14 to 37 to schedule $J$. Recall that policies B and C are used to schedule a small MH job and a large job, respectively. If $J$ is a small MH job, the task scheduler directly inserts $J$'s map tasks to the permanent map-task queue of the determined datacenter (see line 22), and also inserts $J$'s reduce tasks to the permanent reduce-task queue of the determined datacenter (see line 33). In other words, no additional queue will be created for any small jobs. The purpose is not to increase the queue management overhead of JoSS.

In another case, if $J$ is a large job, the task scheduler additionally generates a new map-task queue and a new reduce-task queue to respectively put $J$'s map tasks and $J$'s reduce tasks (see lines 24 to 26 and lines 35 to 37). This will allow the task assigner to properly assign small jobs and large jobs to VPSs.

Recall that two variations of JoSS (i.e., JoSS-T and JoSS-J) are proposed in this study. The former combines the abovementioned task scheduler and a Task-driven Task Assigner (TTA) to provide a fast task assignment. The latter combines the task scheduler and a Job-driven Task Assigner (JTA) to further improve the VPS-locality.

Fig. 5 illustrates how TTA works. Whenever $VPS_{c,\ell}$ has an idle map slot, TTA preferentially assigns a map task from $MQ_{FIFO}$ to $VPS_{c,\ell}$ based on the Hadoop FIFO algorithm (see lines 7 to 8). The goal is to preferentially execute all newly submitted jobs one by one and obtain their filtering-percentage values to determine their job classifications. However, if $MQ_{FIFO}$ is empty, TTA assigns one of the first map tasks from all the other map-task queues of $cen_c$ in a round-robin fashion (see lines 10 to 13) such that tasks can be assigned quickly and job starvation can be avoided.

Similarly, whenever $VPS_{c,\ell}$ has an idle reduce slot, TTA preferentially assigns a reduce task from $RQ_{FIFO}$ to $VPS_{c,\ell}$ (see lines 16 to 17). Only when $RQ_{FIFO}$ is empty, TTA assigns one of the first reduce tasks from other reduce-task queues of $cen_c$ to $VPS_{c,\ell}$ (see lines 19 to 22).

Fig. 6 shows the algorithm of JTA, which in fact is very similar to that of TTA. The only difference is that JTA always uses the Hadoop FIFO algorithm to assign a map task from each map-task queue (please compare line 11 in both Figs. 5 and 6) so as to further improve the VPS-locality.

## 5 SELECTING THE BEST THRESHOLD

Recall that JoSS uses $td$ as a threshold to characterize jobs into RH or MH (see Eq. (3)). In this section, we show how to derive the best value for $td$. To do this, we consider the worst-case inter-datacenter traffic for transmitting the map-input data and reduce-input data of a job when this job, say $J$, is separately judged as a RH job and a MH job.

If $J$ is classified as a RH job, policy A will be used to schedule $J$. The worst case for $J$'s mappers is that all of them need to retrieve their input blocks from other datacenters. However, because of policy A, $J$'s reducers can completely retrieve their input from their local datacenters. Hence, the worst-case inter-datacenter traffic for this classification, denoted by $TR_1$, is

$$TR_1 = \sum_{i=1}^{m} |B_i|. \qquad (5)$$

TABLE 5
The Average Filtering Percentages of Five Benchmarks

| Benchmark | Average filtering percentage |
|---|---|
| WC | 1.039 |
| SC | 0.569 |
| II | 1.166 |
| Grep | 0.10 |
| Permu | 3 |

On the other hand, if $J$ is classified as a MH job, policy B will be used, which guarantees that all mappers of $J$ can always retrieve their input blocks from their local datacenters. But in the worst case (i.e., all map tasks of $J$ are evenly scheduled to each datacenter because of the even distribution of $J$'s input blocks over all datacenters), $J$'s reducers have to retrieve the $\frac{k-1}{k}$ of their input from other datacenters where $k$ is the total number of the datacenters comprising the virtual MapReduce cluster. Hence, the worst-case inter-datacenter traffic for this classification, denoted by $TR_2$, is

$$TR_2 = \frac{k-1}{k} \cdot \sum_{i=1}^{m} |B_i| \cdot FP_J. \qquad (6)$$

If $TR_2 > TR_1$, $J$ should be determined as a RH job (rather than a MH job) since the related worst-case inter-datacenter traffic is less. Otherwise, $J$ should be determined as a MH job, rather than a RH job. In fact, $TR_2 > TR_1$ can be expressed as $\frac{k-1}{k} \cdot \sum_{i=1}^{m} |B_i| \cdot FP_J > \sum_{i=1}^{m} |B_i|$, which also implies that $\frac{k-1}{k} \cdot FP_J > 1$. Hence, we can obtain Eq. (7).

$$FP_J > \frac{k}{k-1} \qquad (7)$$

With Eq. (7) and the condition to determine a RH job (i.e., $FP_J > td$ shown in Eq. (3)), we can derive the best value of $td$, i.e.,

$$td = \frac{k}{k-1}. \qquad (8)$$

## 6   PERFORMANCE EVALUATION AND COMPARISON

In this section, we evaluate and compare JoSS-T and JoSS-J with three scheduling algorithms provided by Hadoop, including the FIFO algorithm (FIFO for short), the fair scheduling algorithm (Fair for short), and the capacity algorithm (Capa for short).

We established a virtual MapReduce cluster by renting 31 VPSs from Linode [12], which is a privately owned VPS provider based in New Jersey. One VPS acts as the Hadoop master server and is located at a datacenter in Dallas. The remaining 30 VPSs act as slaves. Among them, 15 VPSs are located at a datacenter in Dallas and the other 15 VPSs are located at a datacenter in Atlanta. Each VPS runs Ubuntu 10.04 with two CPU cores, 2 GB RAM, and 48 GB SSD storage space. Each VPS has a map slot and a reduce slot. We use Hadoop MRv1, which is widely adopted in production settings [28], as the implementation of MapReduce, and modify the source code of Hadoop-0.20.2 to realize JoSS-T and JoSS-J.

To study how different MapReduce jobs with different filtering-percentage values influence the performances of the five tested algorithms, we chose the following five MapReduce benchmarks to conduct our experiments. The first four jobs are from the MapReduce benchmark suite called PUMA [29], and the corresponding input data are web documents chosen from [30]. The last one job is created by ourselves, and its input data is a set of TXT files chosen from [31]. Based on our analyses shown in Section 4.1, Table 5 lists the average filtering-percentage values of these benchmarks.

TABLE 6
The Details of the Small Workload

| Total number of jobs | 300 |
|---|---|
| # of WC jobs | 60 |
| # of SC jobs | 59 |
| # of II jobs | 59 |
| # of Grep jobs | 61 |
| # of Permu jobs | 61 |
| Average job arrival interval/ Standard deviation | 27.70 sec/ 36.52 sec |

1) Word-Count (WC for short), which counts the occurrence of each word in data files.
2) Sequence-Count (SC for short), which generates a count of all unique sets of three consecutive words in data files.
3) Inverted-Index (II for short), which takes a list of data files as input and generates word-to-file indexing.
4) Grep, which searches for a pattern in data files.
5) Permu, which generates the permutation for three consecutive DNA sequences in DNA data files.

Based on Eq. (8) and our virtual MapReduce cluster, $td = 2$ ($= \frac{k}{k-1} = \frac{2}{1}$). Consequently, not all tested MapReduce benchmarks will be classified as the same job type by JoSS-T and JoSS-J. Some of them will be classified as MH jobs, and the others will be classified as RH jobs.

We used the above five benchmarks to create a small workload and a mixed workload, and used the two workloads to evaluate the performances of the five algorithms. The details are listed in Tables 6 and 7. The small workload consists of 300 jobs. The size of the input data processed by each job is approximately 1 GB. The mixed workload comprises 100 jobs to process input-data ranging from 1 to 12 GB.

TABLE 7
The Details of the Mixed Workload

| Total number of jobs | 100 | |
|---|---|---|
| # of 1 GB jobs | 64 | 26 WC jobs 20 II jobs 10 SC jobs 5 Grep jobs 3 Permu jobs |
| # of 5 GB jobs | 19 | 19 Permu jobs |
| # of 12 GB jobs | 17 | 6 WC jobs 11 II jobs |
| Average job arrival interval/ Standard deviation | 42.26 sec/ 50.13 sec | |

The submission orders of all jobs in two workloads were randomly determined, and they are fixed for all tested algorithms to achieve a fair comparison. Hence, there is only one permutation for each workload, and each tested algorithm separately performed the small workload and the mixed workload once. In the small workload, the inter-arrival intervals of all jobs were generated by a workload synthesis tool called SWIM [27]. By doing so, all jobs of the small workload will be submitted to each tested algorithm one by one based on their submission orders and the generated inter-arrival time. Due to the fact that employing the inter-arrival intervals generated by SWIM for the mixed workload will make Hadoop too busy and unstable, we generated all the inter-arrival time of the mixed workload based on a Poisson distribution [32]. This is the reason why the average job arrival interval of the mixed workload is lager than that of the small workload.

In both workloads, each job has only one reduce task, and all data files were partitioned into blocks of 128 MB. Based on the block size, each job in the small workload has eight ($= 1{,}024/128$) map tasks. Hence, all jobs in this workload must be classified as small jobs by JoSS-T and JoSS-J based on Eq. (4). However, not all jobs in the mixed workload will be classified as small jobs by JoSS-T and JoSS-J. The reason why the two workloads mostly focus on small jobs is because that the survey results in [17] showed that most jobs in Facebook workload traces are small. Hence, we followed the similar idea to design the two workloads and determine the distribution of jobs. In fact, the two workloads have different design goals. The small workload is to evaluate each algorithm when all submitted jobs are either small MH jobs or small RH jobs. However, the mixed workload is to evaluate each algorithm when all submitted jobs are not only small MH jobs and small RH jobs, but also large jobs.

Recall that JoSS-T and JoSS-J are designed for a tenant with a limited budget to improve job performance in his/ her virtual MapReduce cluster. In other words, the storage space of the cluster might be very limited. To see how each tested algorithm performs in such resource-limited computing environment, in our experiments, each data block has only one replica.

## 6.1 The Small Workload

The following metrics are used to evaluate the performance of the five algorithms under the small workload.

1. Map-data locality, which can be divided into VPS-locality rate, Cen-locality rate, and off-Cen rate as shown in Eqs. (9), (10), and (11), respectively.

$$\text{VPS-locality rate} = \frac{\#_{VPS}}{\mathcal{M}}, \tag{9}$$

$$\text{Cen-locality rate} = \frac{\#_{Cen}}{\mathcal{M}}, \tag{10}$$

$$\text{off-Cen rate} = 1 - \left(\frac{\#_{VPS}}{\mathcal{M}} + \frac{\#_{Cen}}{\mathcal{M}}\right), \tag{11}$$

where $\#_{VPS}$ and $\#_{Cen}$ are respectively the total numbers of the map tasks that can achieve the VPS-locality and the Cen-locality, and $\mathcal{M}$ is the total number of the map tasks in the workload. Note that the values of the above three rates range from 0 to 1. A value of one is desirable for both the VPS-locality rate and the Cen-locality rate, but a value of zero is desirable for the off-Cen rate.

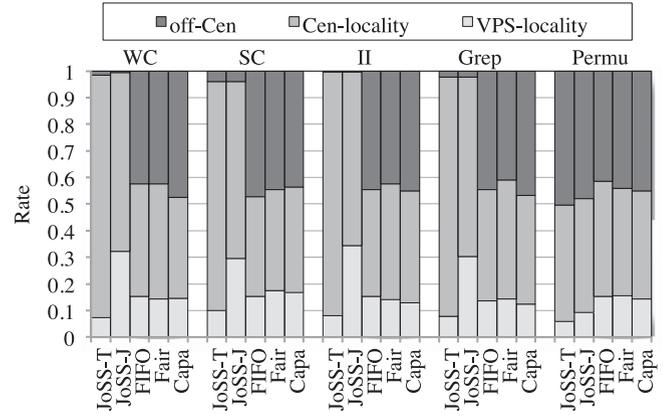

Fig. 7. The map-data locality results of the five tested algorithms under the small workload.

2. Reduce-data locality rate, which is defined as the percentage of input data that a reducer can obtain from its local datacenter. The value ranges from 0 to 1. A value of one is desirable.

3. Inter-datacenter network traffic (INT for short), which is the total inter-datacenter network traffic generated during the execution of the workload. A small value of INT is desirable.

4. Job turnaround time (JTT for short), which starts when a job is submitted to the cluster and finishes when the job is completed. A short JTT is desirable.

5. VPS load, which shows the average number of map tasks executed by each VPS and the corresponding standard deviation. With this metric, we can know the load balance among VPSs. A small standard deviation is desirable.

Fig. 7 illustrates the map-data locality results of all algorithms under the small workload. When JoSS-T and JoSS-J were used to run small MH jobs (i.e., those WC, SC, II, and Grep jobs), the corresponding off-Cen rates are not only far lower than those of the other algorithms, but also close to zero, implying that all the mappers can almost retrieve their input blocks from their local datacenters. This is because policy B (which favors map-data locality) is always used by JoSS-T and JoSS-J to schedule small MH jobs.

However, the above phenomenon did not appear when JoSS-T and JoSS-J performed Permu jobs (which are RH jobs) since policy A (which favors reduce-data locality, rather than map-data locality) is always used to schedule small RH jobs.

Even though JoSS-T and JoSS-J had similar off-Cen result, the latter provided a higher VPS-locality rate since it employs the JTA (see Fig. 6) to further increase the VPS-locality. This property also makes the VPS-locality rate of JoSS-J higher than those of the other algorithms when the executed jobs are small MH jobs.

Fig. 8 illustrates the reduce-data locality results of all algorithms. Since JoSS-T and JoSS-J employ the same reduce-task scheduling approach, they have a very similar reduce-data locality rate in every benchmark. In addition, it is clear that JoSS-T and JoSS-J provided a higher reduce-data locality rate than the other three algorithms, especially when RH jobs were executed. The reason is the same, i.e., JoSS-T and JoSS-J always use policy A (which favors reduce-data locality) to schedule small RH jobs.

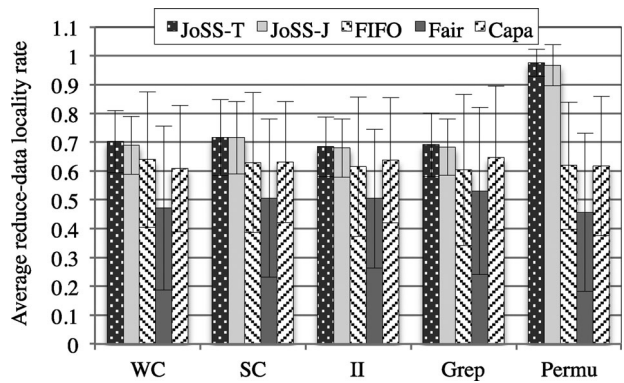

Fig. 8. The average reduce-data locality rates of the five algorithms under the small workload.

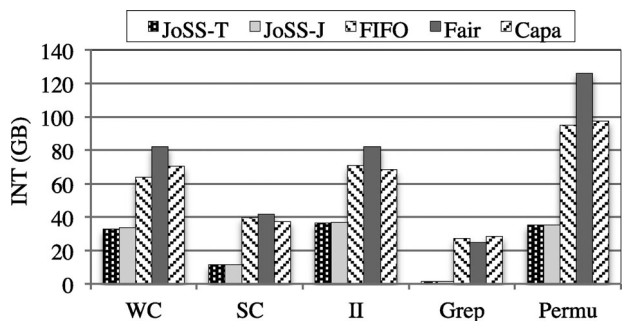

Fig. 9. The INT of the five algorithms on the small workload.

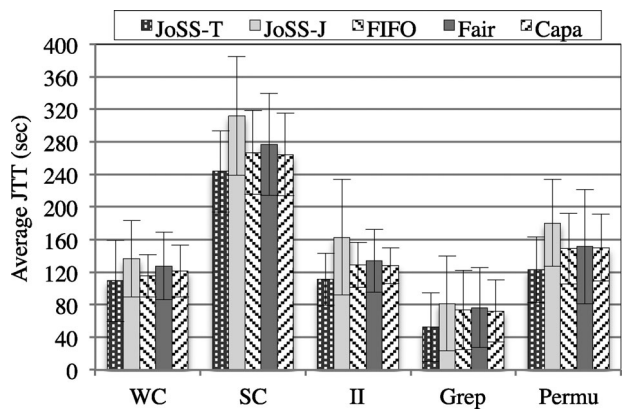

Fig. 10. The average JTT of the five algorithms on the small workload.

Fig. 9 shows that JoSS-T and JoSS-J consumed similar and low inter-datacenter network traffic because they had close off-Cen and reduce-data locality rates, no matter which benchmark was performed. Furthermore, we can see that both JoSS-T and JoSS-J have a much lower INT compared with the other three algorithms since the two algorithms have superior performances in terms of data locality.

Fig. 10 illustrates the average JTT results of all the algorithms under the small workload. No matter which benchmark was executed, JoSS-T led to the shortest average JTT

### TABLE 8
### The Normalized JTT Values of the Five Algorithms

| Algorithm | WC | SC | II | Grep | Permu |
|-----------|------|------|------|------|-------|
| JoSS-T | 1 | 1 | 1 | 1 | 1 |
| JoSS-J | 1.25 | 1.28 | 1.46 | 1.55 | 1.47 |
| FIFO | 1.05 | 1.09 | 1.16 | 1.40 | 1.21 |
| Fair | 1.17 | 1.14 | 1.21 | 1.45 | 1.23 |
| Capa | 1.11 | 1.08 | 1.15 | 1.37 | 1.22 |

### TABLE 9
### The Average VPS Loads when the Five Algorithms Perform the Small Workload

| Algorithm | Average number of tasks executed by each VPS | Standard deviation |
|-----------|-----------|-----------|
| JoSS-T | 80 | 13.58 |
| JoSS-J | 80 | 13.59 |
| FIFO | 80 | 6.32 |
| Fair | 80 | 9.64 |
| Capa | 80 | 9.81 |

among all algorithms. Table 8 further shows the normalized JTT of all algorithms compared with JoSS-T. We can see that JoSS-J caused the longest JTT among all tested algorithms. Recall that the only difference between JoSS-T and JoSS-J is that the former uses TTA (see Fig. 5) to assign tasks, whereas the latter uses JTA (see Fig. 6) to assign tasks. Since JTA uses the Hadoop FIFO algorithm (which follows the strict FIFO order and prefers to schedule map-local tasks first) to further improve VPS-locality, the execution of some map tasks might be delayed, which therefore prolonged the corresponding JTT for JoSS-J.

The results confirm that using JoSS-T to schedule the small workload not only reduces off-Cen rate and improves reduce-data locality, but also shortens the corresponding JTT.

Table 9 lists the average number of map tasks executed by each VPS when the five algorithms individually performed the small workload. Regardless of the tested algorithm, the average number of map tasks performed by each VPS is 80 (= 2,400/30), but it is inevitable that JoSS-T and JoSS-J have a higher standard deviation than the other algorithms because of policies A and B. Among all tested algorithms, FIFO achieved the best load balance since its standard deviation was the smallest, but this advantage did not improve FIFO's performance in terms of JTT.

## 6.2 The Mixed Workload

In this subsection, we evaluated how the five algorithms perform when they execute the mixed workload. Similar to the metrics used earlier, the map-data locality, reduce-data locality, INT, and VPS load were also used to evaluate the five algorithms. However, JTT was not considered in this experiment since the input sizes processed by the jobs in the mixed workload were different, which makes this metric meaningless. Hence, we further used the following metrics to better measure these algorithms:

- Workload turnaround time (WTT for short), which is the total time required by the cluster to complete the entire mixed workload.

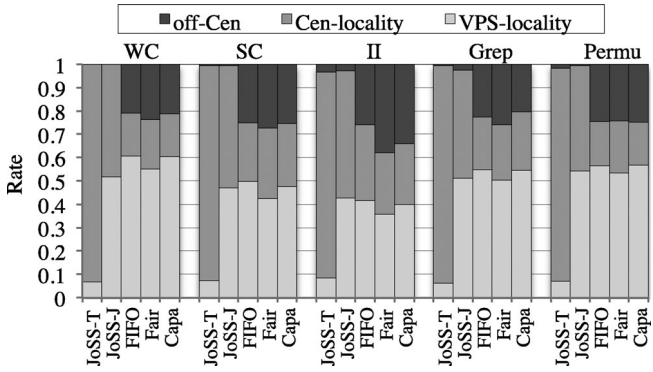

Fig. 11. The map-data locality results of the five tested algorithms on the mixed workload.

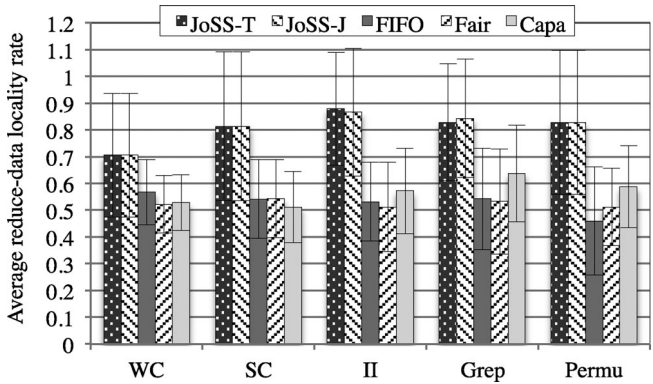

Fig. 12. The average reduce-data locality rates of the five algorithms under the mixed workload.

- Cumulative job completion rate during the execution of the mixed workload.

Fig. 11 illustrates the map-data locality results of all algorithms under the mixed workload. Among all algorithms, JoSS-T caused the lowest VPS-locality rate, regardless of job type. The reason is obvious, i.e., JoSS-T uses TTA to quickly assign a task to an idle VPS, rather than increasing the VPS-locality.

On the other hand, by comparing Figs. 11 and 7, we can see that the VPS-locality rates of the other four algorithms on the mixed workload increased. This is because each VPS held more input blocks of large jobs and therefore improved the VPS-locality rate. This property also causes that JoSS-J was not always better than those of the other three algorithms in terms of VPS-locality. Nevertheless, for all tested MH jobs (i.e., WC, SC, II, and Grep jobs), JoSS-T and JoSS-J had similar off-Cen rates, which were still much lower than those of the other three algorithms.

Fig. 11 also shows that when the executed jobs were classified as RH jobs (i.e., Permu), the off-Cen rates of JoSS-T and JoSS-J were no longer as large as they were in Fig. 7. This is because those large RH jobs in the mixed workload were always scheduled by policy C, which favors map-data locality, especially for the Cen-locality.

Fig. 12 shows the reduce-data locality results of all algorithms under the mixed workload. It is clear that the two variations of JoSS outperform the other algorithms in every benchmark and job type. But we can see that when JoSS-T and JoSS-J executed the RH jobs (i.e., Permu) in the mixed workload, the corresponding reduce-data locality rates were no longer as high as they were in Fig. 8. The reason is the same, i.e., all the large RH jobs were scheduled by policy C.

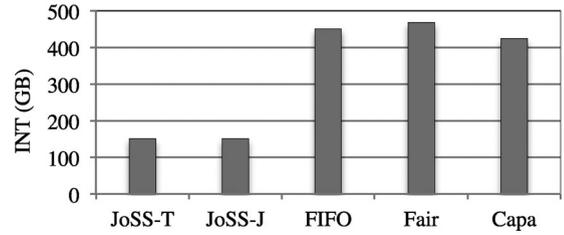

Fig. 13. The INT of the five algorithms on the mixed workload.

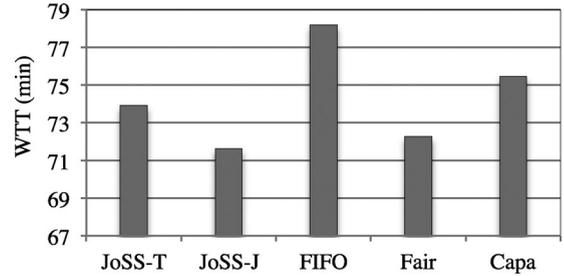

Fig. 14. The WTTs of the five algorithms under the mixed workload.

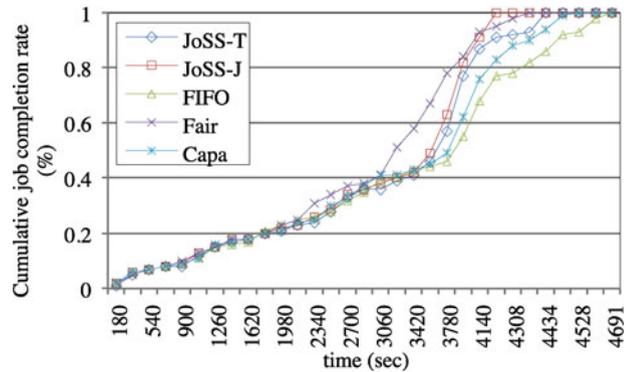

Fig. 15. The cumulative job completion rates of the five algorithms under the mixed workload.

Since JoSS-T and JoSS-J had good data-locality performances (see Figs. 11 and 12), they dramatically reduced the inter-datacenter network traffic for retrieving map-input data and reduce-input data during the execution of the mixed workload. The results depicted in Fig. 13 shows that the INTs of JoSS-T and JoSS-J are only 33.44, 32.16, 35.43 percent of those of FIFO, Fair, and Capa, respectively.

Fig. 14 illustrates the WTT results of the five algorithms on the mixed workload. Among all algorithms, FIFO consumed the longest time to complete the entire mixed workload. This is because some larger jobs arrived to the cluster first, smaller jobs afterwards. Thus, smaller jobs were delayed and long JTT occurred, which therefore prolonged WTT.

On the contrary, JoSS-J led to the shortest WTT among all tested algorithms since it not only used policies A, B, C to respectively schedule small MH jobs, small RH jobs, and large jobs, but also employed JTA to schedule all jobs in a round-robin fashion and meanwhile achieve VPS-locality. Although JoSS-T also followed the three policies to schedule tasks, its WTT performance was not as good as JoSS-J's, implying that employing TTA to assign tasks cannot

TABLE 10
The Average VPS Loads when the Five Algorithms Were
Individually Used to Perform the Mixed Workload

| Algorithm | Average number of tasks run by each VPS | Standard deviation |
|---|---|---|
| JoSS-T | 98.23 | 7.78 |
| JoSS-J | 98.23 | 11.06 |
| FIFO | 98.23 | 18.30 |
| Fair | 98.23 | 9.46 |
| Capa | 98.23 | 14.74 |

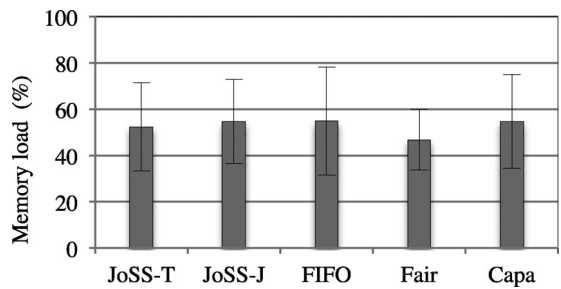

Fig. 17. The Memory load of the Hadoop master server when the five algorithms are individually used to execute the mixed workload.

effectively shorten WTT when a MapReduce workload includes large jobs.

Fig. 15 shows the corresponding cumulative job completion rates of all algorithms. We can see that the Fair algorithm performed best before the mixed workload was executed for 4,140 seconds. However, after that, JoSS-J completed the entire mixed workload first.

Table 10 lists the average VPS load when the five algorithms were individually used to execute the mixed workload. Among all tested algorithms, JoSS-T led to the smallest standard deviation (i.e., the best load balance between VPSs), which is much lower than its standard deviation shown in Table 9. The main reason is that JoSS-T uses policy C to schedule all large jobs of the mixed workload and uses TTA to quickly assign each head-of-queue task to an idle VPS. Similarly, employing policy C to schedule large jobs also improves the load balance of JoSS-J (Please compare Table 10 with Table 9), even though the load balance of JoSS-J shown in Table 10 is still no better than that of Fair.

### 6.3 Scheduling Overhead

In this subsection, we evaluate the overhead caused by each tested algorithm. Figs. 16 and 17 respectively show the CPU idle rate and memory load of the Hadoop master server when the five algorithms separately executed the mixed workload. It is clear that both JoSS-T and JoSS-J did not significantly increase the CPU and memory load of the master server compared with the other algorithms.

In addition, we further evaluated the extra storage space consumed by JoSS-T and JoSS-J to store all necessary information about every newly executed job, including the corresponding hash value and average filtering-percentage value. In our experiments, each such a record is about 20 bytes. Hence, the total storage consumption is proportional to the number of the newly executed jobs.

Based on the above analyses, it is clear that JoSS-T and JoSS-J do not incur significant computation overhead,

memory overhead, and storage overhead to the Hadoop master server.

## 7 CONCLUSION AND FUTURE WORK

In this paper, we have introduced JoSS for scheduling MapReduce jobs in a virtual MapReduce cluster consisting of a set of VPSs rented from a VPS provider. Different from current MapReduce scheduling algorithms, JoSS takes both the map-data locality and reduce-data locality of a virtual MapReduce cluster into consideration. JoSS classifies jobs into three job types, i.e., small map-heavy job, small reduce-heavy job, and large job, and introduced appropriate policies to schedule each type of job. In addition, the two variations of JoSS (i.e., JoSS-T and JoSS-J) are further introduced to respectively achieve a fast task assignment and improve the VPS-locality.

The extensive experimental results demonstrate that both JoSS-T and JoSS-J provide a better map-data locality, achieve a higher reduce-data locality, and cause much less inter-datacenter network traffic as compared with current scheduling algorithms employed by Hadoop. The experimental results also show that when the jobs of a MapReduce workload are all small to the underlying virtual MapReduce cluster, employing JoSS-T is more suitable than the other algorithms since JoSS-T provides the shortest job turnaround time. On the other hand, when the jobs of a MapReduce workload are not all small to the virtual MapReduce cluster, adopting JoSS-J is more appropriate because it leads to the shortest workload turnaround time. In addition, the two variations of JoSS have a comparable load balance and do not impose a significant overhead on the Hadoop master server compared with the other algorithms.

In the future, we would like to extend JoSS by taking heterogeneous virtual MapReduce clusters into consideration so as to increase the flexibility of JoSS.

### ACKNOWLEDGMENTS

The authors thank the scholarship of the Sandwich Programme supported by Ministry of Science and Technology, Taiwan and Deutscher Akademischer Austausch Dienst (DAAD) under Grants NSC 102-2911-I-100-524 and NSC 101-2911-I-009-020-2. J.-C. Lin is the corresponding author.

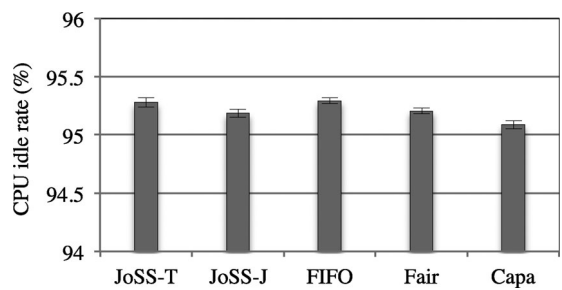

Fig. 16. The CPU idle rate of the Hadoop master server when the five algorithms are individually used to execute the mixed workload.

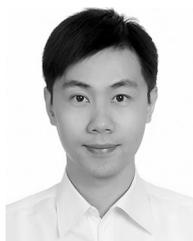

**Ming-Chang Lee** received his MS degree in Computer Science Department from TungHai University, Taiwan, in 2006. In March 2015, he received his PhD degree from Department of Computer Science, National Chiao Tung University, Taiwan. During his PhD program, he received a scholarship of NSC-DAAD Sandwich Program in 2012 and did research with Prof. Ramin Yahyapour at University of Göttingen, Germany. His research interests are in the field of evaluating and improving job turnaround time, reliability, data availability, and energy consumption of distributed and parallel systems, including cloud computing, grid computing, data grid, MapReduce, and YouTube. He is also interested in the field of sentiment analysis on Big Data.

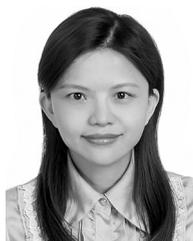

**Jia-Chun Lin** received her MS degree in computer science from TungHai University, Taiwan, in 2005. She received a scholarship of NSC-DAAD Sandwich Program in 2013 and conducted a research with Prof. Ramin Yahyapour at the University of Göttingen, Germany, from March 2014 to February 2015. In March 2015, she received her PhD degree in computer science and engineering from National Chiao Tung University, Taiwan. She is currently a postdoctoral research fellow at University of Oslo, Norway. Her research interests include distributed and parallel computing, cloud computing, job scheduling, reliability analysis, energy consumption, and scalability issues in MapReduce framework. She is also interested in software product line, software verification and validation, and formal methods.

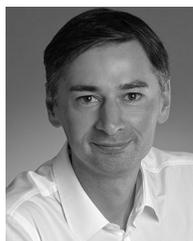

**Ramin Yahyapour** is a full professor at the Georg-August University of Göttingen since October 2011. He is also the managing director in the GWDG, a joint compute and IT competence center of the university and the Max Planck Society. He is also the CIO of the University and the University Medical Center Göttingen (UMG). Before his appointment in Göttingen, he was a professor at TU Dortmund University, the director of the IT & Media Center and CIO of the University. His research interest lies in the area of efficient resource management in its application to service-oriented infrastructures, clouds, and data management. He is especially interested in data and computing services for eScience.

▷ **For more information on this or any other computing topic, please visit our Digital Library at** www.computer.org/publications/dlib.